# Persistent enhancement of the carrier density in electron irradiated InAs nanowires

Corentin Durand, Maxime Berthe, Younes Makoudi, Jean-Philippe Nys, Renaud Leturcq, Philippe Caroff, Bruno Grandidier*

*Institut d'Electronique, de Microélectronique et de Nanotechnologie, IEMN, (CNRS, UMR 8520, Département ISEN, 41 bd Vauban, 59046 Lille Cedex, France.*

**ABSTRACT**

We report a significant and persistent enhancement of the conductivity in free-standing non intentionnaly doped InAs nanowires upon irradiation in ultra high vacuum. Combining four-point probe transport measurements performed on nanowires with different surface chemistries, field-effect based measurements and numerical simulations of the electron density, the change of the conductivity is found to be caused by the increase of the surface free carrier concentration. Although an electron beam of a few keV, typically used for the inspection and the processing of materials, propagates through the entire nanowire cross-section, we demonstrate that the nanowire electrical properties are predominantly affected by radiation-induced defects occuring at the nanowire surface and not in the bulk.

* E-mail: bruno.grandidier@isen.iemn.univ-lille1.fr

# 1. Introduction

Semiconductor devices are more and more relying on the integration of nanostructures to enhance their optical and electronic performances. In order to comply with a high level of quality and reliability, the controlled positioning of the nanostructures in the devices and the subsequent inspection of their integrity during the manufacturing processes require the use of high resolution analytical and imaging techniques. Devices consisting of single semiconductor nanowires (NWs) are generally not an exception to this rule. For example, the inspection, localization and connection of nanowires with electrical electrodes are hard to achieve without an electron beam [1]. However, the excitation of materials with incident electrons can stimulate a wealth of reactions that are known to modify the material properties [2,3]. For years, the tolerance of materials and devices to a beam of electrons has indeed been of great concern in many technological applications, because electrons are known to produce defects and generate disorder in materials [4,5,6].

While outstanding performances or physical effects have been recently demonstrated with semiconductor NWs [7,8,9,10,11], a detailed study of the electrical modifications that are caused by electron beam irradiation is still missing. For example, contact processing or contact through a retrofitted probe are usually achieved on InAs NWs with the help of a scanning electron microscope (SEM) working with an energy beam in the range of 5 to 20 keV [12,13]. Within this energy range, the electrons interact with the entire section of the NWs [14]. As it is well established that electron irradiation of InAs bulk material causes a Fermi level pinning high in the conduction band [15], questions arise to determine to what extend an electron beam might change the electrical properties of InAs NWs.

A serious impediment to study the direct impact of electron irradiation on the NW conductivity resides in the presence of remote charges in the NW environment, that can electrostatically control their conductivity [16]. In order to circumvent this drawback, sophisticated characterisation techniques must be developed, based, for example, on multiple probe scanning tunneling microscopy (STM), that enable arbitrary arrangements of probe electrodes on one-dimensional nanostructures [17,18]. Taking advantage of this technique, we therefore investigate electrical transport in non intentionnaly doped InAs NWs that were subject to electron irradiation. Upon irradiation, we find a significant and reproducible decrease of the resistivity, leading to NWs with a metallic-like character. This phenomenon is persistent in ultra high vacuum (UHV) and the NWs barely recover their initial resistivity once the NWs have been exposed to air. While the transport measurements show a degradation of the mobility with the irradiation, this effect is largely counterbalanced by a strong increase of the electron concentration. Study of the NW resistivity dependence on the surface chemistry provides a compelling evidence for a conduction dominated by accumulated electrons. Such a localized increase of the electron density in the NWs, that is supported by numerical analysis, allows discriminating between the contribution of surface and bulk charged defects.

## 2. Methods

[0001]-oriented wurtzite InAs NWs were grown from gold particle seeds using gas source molecular beam epitaxy. Starting with a short InP stem grown at 410°C on InP(111)B substrates, InAs NWs were subsequently obtained by direct switch from $P_2$ to $As_2$ molecular fluxes. The NWs are of very high crystallographic quality, with only a very limited number of single stacking faults along their growth axis, as shown previously [19]. Two types of geometry were electrically investigated: interconnected NWs with conventional planar technology and free-standing NWs. For the first ones, NWs were deposited onto a degenerately-doped Si wafer with a 225 nm of thermally-grown $SiO_2$ layer and selected for contact processing with a scanning electron microscope at low magnification and with a short integration time. After spin-coating the sample with an electron-beam sensitive polymer, four contacts are defined along the nanowire using electron beam exposure. The oxide on the contact area was etched in a $(NH_4)_2S_x$ solution just before evaporation of Ti/Au electrodes (10/150 nm thick) [20]. This step was followed by a thorough lift-off to get NWs with clean oxidized sidewalls.

As to the second type of geometry, the as-grown NWs were kept on the InP substrate. The NW mean diameter was 70 ± 9 nm, but NWs with smaller and bigger diameters could sporadically be produced. All the experiments were carried out with the same setup, consisting of a SEM and a multiprobe STM (Nanoprobe, Omicron Nanotechnology), that was operated with a control system able to run the four STM tips independently (Nanonis multiprobe control system, SPECS). The electrical measurements were performed in UHV (base pressure lower than $5 \times 10^{-10}$ mbar) and in the dark, always with the electron beam turned off.

In order to investigate how the surface chemistry modifies the Fermi level pinning, three different surface treatments were applied. The NWs were either left in air for several days to ensure a complete oxidation of their sidewalls. Or, they were passivated with an ammonium sulfide solution to produce a chemisorbed S layer on their surface and, then, quickly loaded into the Nanoprobe system. Indeed this chemical strategy has been shown to efficiently remove the native oxide on the InAs (001) surfaces, offering greater chemical stability to electronic devices [21]. Finally, NWs were also capped with a thin layer of As at the end of the growth and subsequently transferred to the Nanoprobe system. The evaporation of the capping layer was achieved at 350°C, monitoring the As desorption with mass spectrometry. This preparation was always achieved prior to the SEM inspection of the NWs and their subsequent electrical characterization. Apart from the existence of small Au clusters, that are caused by the gold diffusion along the NW sidewalls from the seed particle during the evaporation of the As capping layer, such a method allows to restore the initial crystalline surface structure of the NWs [22].

## 3. Results and discussion

The resistance of the NWs and its change with electron irradiation were first investigated for interconnected InAs NWs, with native-oxide covered sidewalls. A CCD camera was used to position the STM tips on metallic pads, thus avoiding the exposure of the NWs to the e-beam, as shown in Figure 1a and b. Two-terminal current-voltage characteristics are found to be linear, and clearly reveal a decrease of the resistance once an InAs NW have been exposed to a 5 keV electron beam for 0.35 s (Figure 1c), in agreement with the litterature [23]. Periodic measurements of the resistivity over 48 hours (events 1 to 5) show that the NW retains its low resistivity after irradiation (Figure 1d). A subsequent exposure to air for four hours allows us to partially restore a high resistivity (event 6). But, as soon as the NW has been again exposed to the electron beam, the resistivity returns to its low value (event 7).

As we can not rule out that the change of the resistivity could be caused by charges trapped in the surroundings of the NWs, since dielectric materials are known to efficiently retain charges in UHV [24], similar free-standing InAs NWs with native-oxide covered sidewalls were characterized by the four-probe method. In such a configuration, four STM tips were brought into electrical contacts with the NW using the distance regulation of the STM control system under the monitoring of the SEM (Figure 2a). Tips 1 and 3 were connected to the seed alloy particle and the base of the NW, respectively. They were used as source and ground electrodes for driving the current through the NW, while the substrate was kept floating. Tips 2 and 4 were the potential probes. In most cases, the I(V) curves between a single tip and the substrate, held at ground, show an ohmic behaviour (see lower inset of Figure 2b).

When a current is passed through a NW, the potential drop measured between the inner probes is found to be linearly proportional to the current intensity (Figure 2b). The gradient of the curves yields the four-probe resistance of the NWs for a given separation between the inner probes. By moving the inner tips towards the center of the NW, a set of $V(I)$ characteristics was acquired for different tip separations $L$. From these $V(I)$ curves, we obtain the resistance $R_{NW}$ that is found to linearly increase with $L$, as shown in the upper inset of Figure 2b. The slope yields a resistance per unit length $R_{NW}/L$ of 6.3 kΩ/μm.

We performed similar measurements on an ensemble of free-standing NWs, that were all subject to irradiation, since their connection with the STM tips can not be managed without the use of the SEM. We also measured the resistance per unit length of interconnected NWs, that were not irradiated. The resistances per unit length of all the NWs were then compared. Figure 3a clearly reveals a decrease of the resistance, when the NWs have undergone an electron irradiation. Such a behaviour is consistent with the one observed in the two-terminal experiments. It unequivocally demonstrates that the increase of conductivity is caused by the action of the incident electrons impinging on the NW. The high conductivity of the irradiated NWs is further evidenced by four-probe measurements performed at low temperatures (Figure 3b). For these experiments, the data points shown in Figure 3 at different temperatures are usually measured on different nanowires, due to the difficulty to keep the tips connected to the same nanowire while changing the temperature. Remarkably, the variation of the resistance per unit length shows no particular sign of dependence on the temperature, indicating a metallic nature for the irradiated NWs. This is in clear contrast with the typical semiconductor behavior found for a non-irradiated NW (inset of Figure 3b), where a three-fold variation is measured for $R_{NW}/L$ in the temperature range 300K-120K.

In order to quantify the carrier concentration in the conduction band, the field effect mobilities were estimated from gate sweeps of the source-drain current $I_{ds}$ at a fixed source-drain voltage $V_{ds}$ of 50 mV (Figure 4). The unmodified InAs NWs exhibit a typical n-type response with increased current at more positive back-gate voltage $V_g$ (spectrum s1). Conversely, upon irradiation, the NWs can not longer be pinched off at $V_g$ higher than -20V (spectrum s2), consistent with the strong metallicity of the irradiated NWs deduced from the four-probe measurements. Then, after air exposure (spectrum s3), the resistance increases again, but the transconductance ($g = dI_{ds}/dV_g$), that is related to the field effect mobility [25], gets worse.

As the change of the transconductance reveals a decrease of the mobility after irradiation, the occurence of a low resistance in the interconnected NWs as well as in the free-standing NWs is the signature of an increase of the electron concentration to compensate for the charges trapped at the NW surface or in the NW. Assuming a diffusive transport [26] and taking the value of the mobility measured after air exposure as an upper bound, we estimated the carrier concentration from the inverse of the product between the elementary charge, the measured mobility and resistivity. For the NW with a diameter of 40 nm, we find an electron concentration of $1.2 \times 10^{19}$ cm$^{-3}$ (Figure 4a). As to the second NW shown in Figure 4b, which has a smaller diameter, the estimated carrier concentration is even larger, reaching $1.4 \times 10^{19}$ cm$^{-3}$. Such carrier concentrations demonstrate a stabilization of the Fermi level quite high in the conduction band.

Although this carrier concentration is in agreement with the huge electron concentration found in proton-irradiated bulk InAs ($1.8 \times 10^{19}$ cm$^{-3}$) [15], InAs surfaces have been shown to exhibit a surface electron accumulation layer that can be quite deep depending on the surface chemistry [27]. In order to determine whether the Fermi level pinning arises from the formation of surface or bulk charged defects, the dependence of the NW resistivity on the surface chemistry was also investigated. In case of a stabilization of the Fermi level due to bulk charged defects, we would expect negligible modifications of the NW resistivity with the surface chemistry. As evidenced in Figure 3a, a change of the surface chemistry for irradiated NWs with similar diameters causes a significant variation of their resistance per unit length. The highest resistance is obtained for the NWs with clean and crystalline sidewalls, whereas the S-passivated NWs are the most conductive NWs. Remarkably, such a trend is consistent with the depth of the accumulation layers measured for non-irradiated InAs surfaces prepared with different chemical treatments [28].

Interestingly, it has been shown, theoretically and experimentally, that In adatoms on different InAs surfaces act as donors and thereby cause the formation of a charge accumulation at the surface [29,30]. As electron irradiation of the GaAs(110) surface produces vacancies consisting of Ga-As pairs [31] and, Langmuir evaporation of anions on similar surfaces is known to occur at room temperature [32], we suspect the irradiation of the InAs sidewalls to substancially increase the initial concentration of isolated In adatoms on the clean and cristalline sidewalls of the bare InAs NWs. Assuming the binding of an In adatom to two As atoms and one In atom, the electron counting rule yields an occupied In dangling bond. As the related state is resonant with the conduction band [29], the formation of In adatoms causes a partial filling of the conduction band, resulting in the creation of positively charged defects on the sidewalls of the NWs. Despite the scarcity of data for the oxidized and S-passivated irradiated InAs surfaces, the desorption of anions and the formation of cation adatoms are also very likely on these surfaces under irradiation and could lead to a similar surface accumulation layer, depending on the atomic configuration of the In adatoms.

In order to get insight into the electron distribution along the radial axis of the NWs, one-dimensional numerial simulations have been performed [33]. To model the numerous interface states on the NW sidewalls, we therefore assumed a positive charge density $n_i$ at the NW surface and varied it in the range $10^{11}$ to $10^{15}$ cm$^{-2}$ (surface atomic density of InAs{10-10}) to account for the generation of charged defects by the electron beam. Figure 5 shows the calculated variation of the spatially average electron concentration $n_{av}$ with $n_i$ for InAs NWs with a diameter of 26 and 40 nm [34]. For values of $n_i$ in the range of $10^{12}$ cm$^{-2}$, that are typical of surface/interface state densities obtained for non-irradiated InAs [23,35], we find $n_{av}$ consistent with the measured electron concentrations, that are marked in Fig. 5 by the horizontal dashed segments at positions s1. After irradiation, $n_i$ increases and causes an increase of $n_{av}$ to fulfill the condition of charge neutrality in the NW (dashed segments s2). Interestingly, the resulting electron concentration is still larger for the smaller diameter, in agreement with charges accumulated at the surface, while bulk charge accumulation would lead to the opposite trend since quantum confinement implies a lower density of states for smaller diameter. In UHV, our estimation of the electron concentration indicates an interface charge density slightly higher than $10^{13}$ cm$^{-2}$. Then, the release of the NWs to the air reduces $n_i$ (dashed segments s3). Such effect is likely related to the partial annihilation of charges trapped on the NW sidewalls in air, while a structural disorder induced by irradiation still exists and causes the observed decrease of electron mobility.

More remarkably, an increase of the electron concentration after irradiation takes along with a narrowing of the accumulation layer, as shown in the inset of Figure 5. In this case, transport mainly occurs close to the NW surface, where the surface charged defects seriously scatters the charge carriers and reduce their mobility. Such a result is consistent with a surface-induced change of the resistance observed for InAs NWs with different surface chemistries. It indicates that a change of the surface state density by impinging energetic electrons is sufficient to explain the huge increase of the electron concentration in the NWs. Therefore the Fermi level in the NWs is not stabilized by charged defects trapped in the bulk of the NWs, but it stays pinned at the NW surface after irradiation. Finally, we would like to point out that the observed decrease of the resistivity under electron irradiation holds for InAs, because the stabilization of the Fermi level in this material occurs far above the bottom of the conduction band [36]. This level is reached when the concentration of defects becomes very high, causing the Fermi level to be pinned by the fulfillment of the condition for charge neutrality. For most of the other semiconductor materials, the charge neutrality level resides in the band gap [36]. The generation of defects thus leaves these materials with a semiconductor character, in agreement with the electrical behaviour recently found for a few irradiated semiconductor nanostructures [37,38].

## 4. Conclusions

In conclusion, we have shown that electron irradiation can significantly change the electrical properties of InAs NWs. At 5 keV, the material modifications are predominantly restricted to the surface of the NWs, involving a deepening of the electron surface accumulation to compensate the surface charged defects. We infer that surface charged defects induced by electron irradiation should occur in all types of semiconductor nanomaterials, causing more or less noticeable changes in their transport properties as a function of the defect concentrations, energy levels and charge states. Such a generation of defects in semiconductor nanostructures could be helpful to guide our understanding of their electronic properties, when they are interfaced with other materials, for example in the creation of heterostructures and metal-semiconductor contacts. As to InAs material, the defect-induced electrostatic effect is significant and pleads for the need of a protective layer. For example, InAs NWs with oxidized or sulfur-passivated sidewalls could be protected with a thin shell of amorphous As material, as the one used to cap the bare InAs NWs with clean and crystalline sidewalls. After the SEM inspection and probe positioning, the protective As capping layer could be desorbed by direct current heating, leaving the surface free of radiation-induced defects.

**Acknowledgments**

This study was financially supported by the EQUIPEX program Excelsior, and C. D. Y.M., R.L., P.C. respectively acknowledge the financial support of the DGA, CEA-LETI and ANR through the project number ANR-11-JS04-002-01. We thank T. Melin, K. Smaali and X. Wallart for discussions.

FIGURE CAPTIONS

**Figure 1.** (a) Large scale SEM image of the sample geometry used to characterize the electrical transport in interconnected InAs nanowires. Two STM tips are plugged to the source and drain electrodes (scale bar: 300 µm). (b) SEM enlarged view of the InAs NW in electrical contact with four electrodes (scale bar: 1µm). (c) Two-terminal current-voltage characteristics of an InAs NW with a diameter of 40 nm, measured before and after a 5 keV irradiation for a dose of $9.2 \times 10^{17}$ e.cm$^{-2}$. (d) NW resistance measured as a function of the treatments, that are described in the main text, for a period of time of 72 hours.

**Figure 2.** (a) SEM image of free-standing InAs NWs grown on an InP(111)B substrate. The NW in the centre is contacted with four W tips, that are used either as sourcing electrodes or potential probes. Electron beam energy: 5 keV. Scale bar: 150 nm. (b) *V-I* curves measured with the four probe method for different spacings *L* between tips 2 and 4. The length *L* is given to the right of the graph. Upper inset: Resistance measured from the *V-I* measurements as a function of *L*. Lower inset: *I-V* curve measured between tip 2 and the substrate.

**Figure 3.** (a) Resistance per unit length $R_{NW}/L$ measured at room temperature as a function of the InAs NW diameter for interconnected InAs NWs (filled symbols) and free-standing and irradiated InAs NWs (open symbols). (b) Temperature dependence of $R_{NW}/L$ for irradiated NWs. Inset: Temperature dependence of $R_{NW}/L$ for a non-irradiated NW with a diameter of 100 nm.

**Figure 4.** Source-drain current versus gate bias measured in ultra high vacuum for two InAs NWs, with a diameter *D*, as a function of the treatment: s1) before irradiation, s2) after an irradiation at 5 keV and doses of a) $9.2 \times 10^{17}$ e.cm$^{-2}$ and b) $4.6 \times 10^{17}$ e.cm$^{-2}$, s3) after an air exposure of 4 hours.

**Figure 5.** Simulated average electron concentration $n_{av}$ as a function of the density of surface states $n_i$ for two InAs slabs with thickness *D* of 26 and 40 nm respectively. The measured

electron concentrations estimated from Fig. 4 for the different treatments s1 to s3 are drawn with dashed segments. Inset: Distribution of the electron density in the slab with a thickness of 40 nm for $n_i=10^{12}$ cm$^{-2}$ (left) and $2\times10^{13}$ cm$^{-2}$ (right).

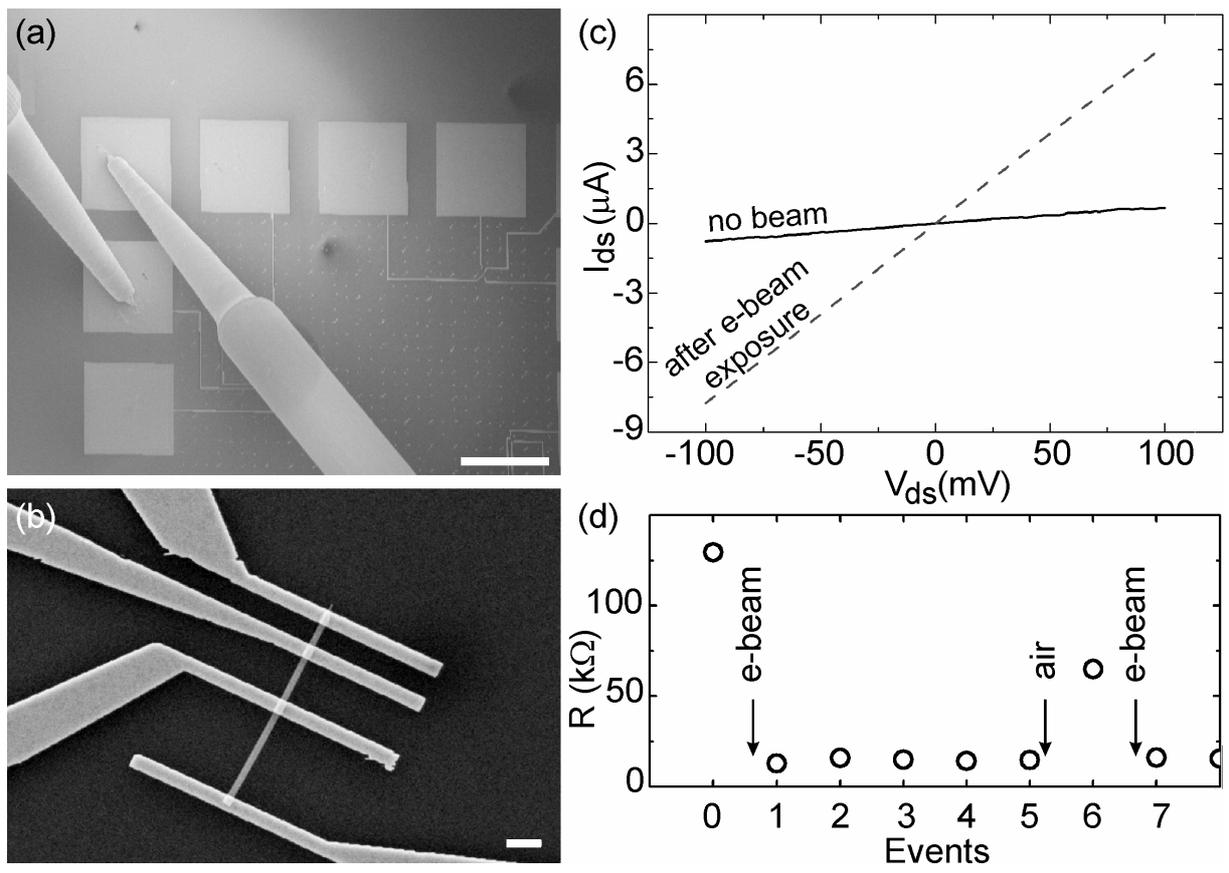

Figure 1

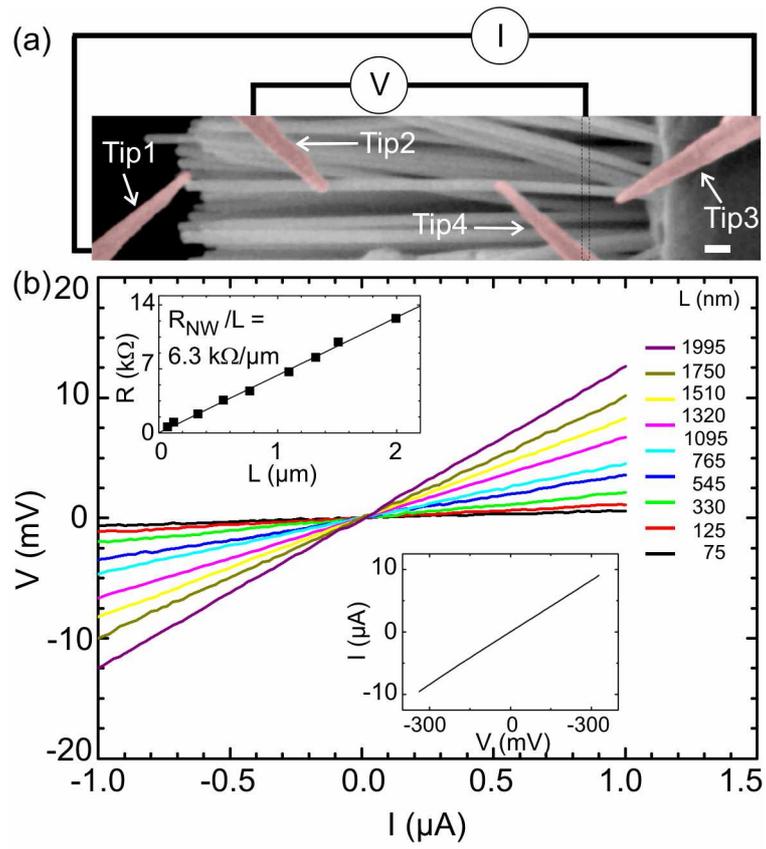

Figure 2

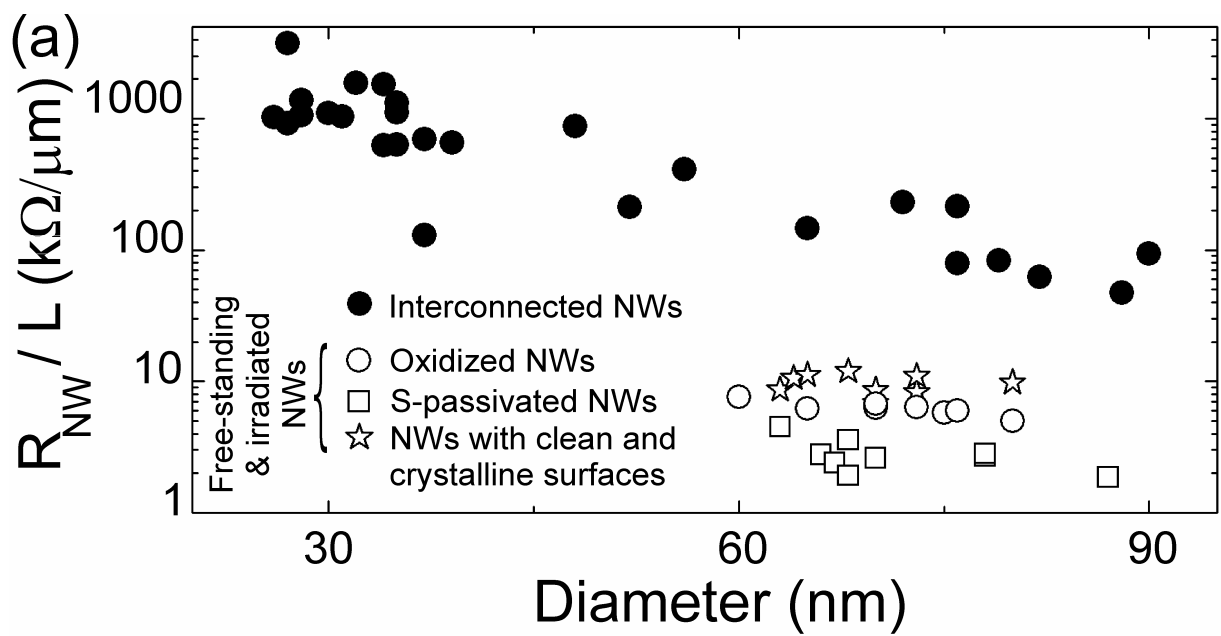

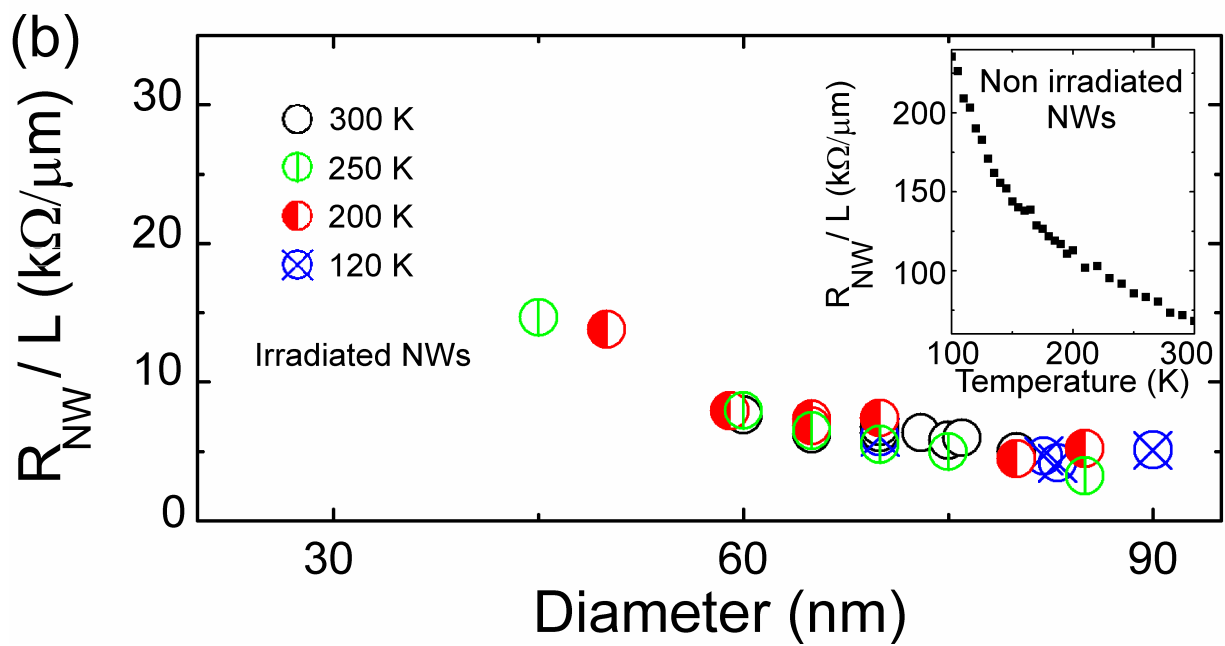

Figure 3

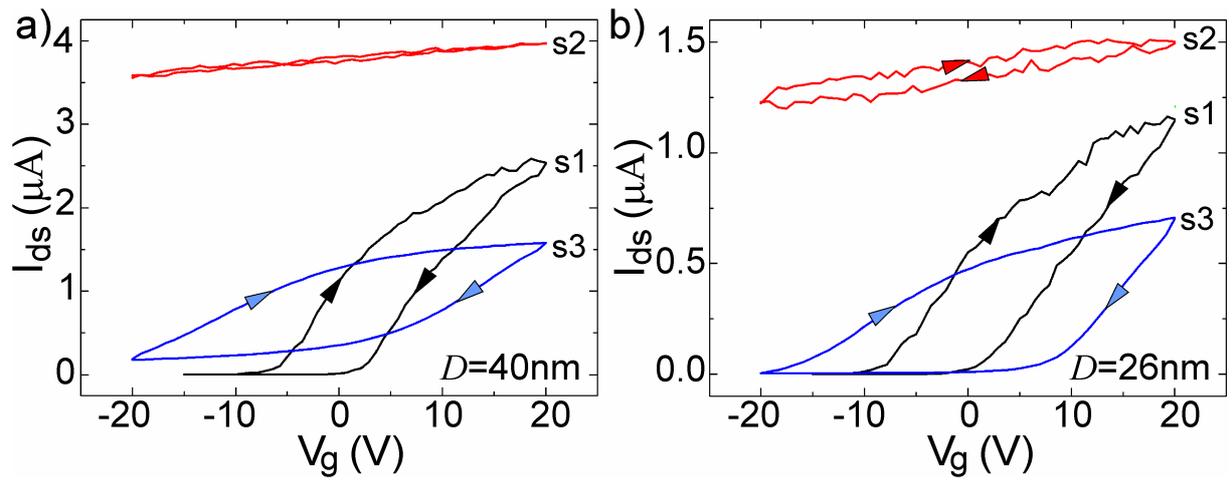

Figure 4

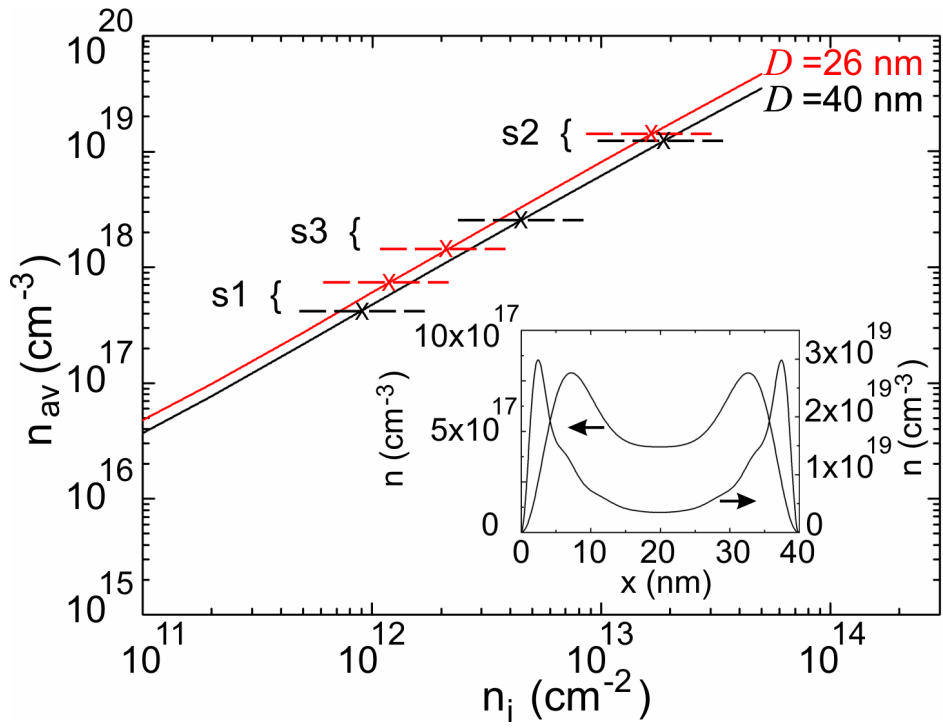

Figure 5